\documentstyle[12pt]{article}
\def\beq{\begin{equation}}
\def\eeq{\end{equation}}
\def\bea{\begin{eqnarray}}
\def\eea{\end{eqnarray}}
\def\bq{\begin{quote}}
\def\eq{\end{quote}}

\def\PL{{\it Phys.Lett.} }

\def\gappeq{\mathrel{\rlap {\raise.5ex\hbox{$>$}}
{\lower.5ex\hbox{$\sim$}}}}

\def\lappeq{\mathrel{\rlap{\raise.5ex\hbox{$<$}}
{\lower.5ex\hbox{$\sim$}}}}

\begin{document}
\pagestyle{empty}
\begin{flushright}
{CERN-TH.96/01}
\end{flushright}
\vspace*{3mm}
\begin{center}
{\bf ASYMPTOTIC PROPERTIES OF THE SOLUTIONS OF }\\
{\bf A DIFFERENTIAL EQUATION APPEARING IN QCD} \\
\vspace*{0.5cm}
{\bf K. Chadan} \\
\vspace{0.1cm}
Laboratoire de Physique Th\'eorique et Hautes
Energies\footnote{Laboratoire associ\'e au Centre
National de la Recherche Scientifique URA D0063.} \\
\vspace*{0.3cm}
{\bf A. Martin} \\
\vspace*{0.1cm}
Theoretical Physics Division, CERN \\
CH - 1211 Geneva 23 \\
and\\
Laboratoire de Physique Th\'eorique ENSLAPP\footnote{URA 14-36 du CNRS
associ\'ee \`a l'Ecole
Normale Sup\'erieure de Lyon et \`a l'Universit\'e de Savoie.} \\
\vspace*{0.1cm}
and\\
\vspace*{0.1cm}
{\bf J. Stubbe}\\
\vspace*{0.1cm}
Institut f\"ur Theoretische Physik - ETH \\
H\"onggerberg, CH - 8093 Z\"urich  \\

\vspace*{0.3cm}
{\bf ABSTRACT} \\
\end{center}
\vspace*{0.1mm}
\noindent
We establish the asymptotic behaviour of the ratio $h^\prime(0)/h(0)$
for
$\lambda\rightarrow\infty$, where $h(r)$ is a solution, vanishing at
infinity, of the differential
equation $h^{\prime\prime}(r) = i\lambda \omega (r) h(r)$ on the domain
$0 \leq r <\infty$ and
$\omega (r) = (1-\sqrt{r} K_1(\sqrt{r}))/r$. Some results are valid for
more general $\omega$'s.

\vspace*{5cm}

\begin{flushleft}
CERN-TH.96/01 \\
January 1996
\end{flushleft}
\vfill\eject

\setcounter{page}{1}
\pagestyle{plain}

\section{Introduction}

Recently \cite{aaa}, \cite{bb} the problem of induced gluon (photon)
radiation in a QCD (QED)
medium has been studied. The authors have been led to study the solution
of a second order
differential equation, which, with some change of notations, reads
\beq
h^{\prime\prime}(r) = i \lambda\omega (r)~ h(r) ~,
\label{1}
\eeq
on the domain $0 \leq r \leq\infty$, with $h(\infty ) = 0$, for large
$\lambda$, for a particular
form of $\omega$ \beq
\omega (r) = {1 - \sqrt{r} K_1 (\sqrt{r}) \over r}~,
\label{2}
\eeq
where $K_1$ is the modified Bessel function of order 1.

Specifically they want to know the asymptotic behaviour of
$h^\prime (0) /h(0)$ for large $\lambda~, \quad h(r)$
being a solution of (\ref{1}) vanishing at infinity. What we want to do
here is to obtain this
limiting behaviour for the specific $\omega$ given by Eq. (\ref{2}) and,
at the same time, study
a broader class of equations of type (\ref{1}).

\section{General Considerations}

$\omega (r)$, as given by Eq. (\ref{2}) has the integral representation
\cite{aaa}:
\beq
\omega (r) = {1 - \sqrt{r} K_1 (\sqrt{r}) \over r } = \int^\infty_0
{2\sin^2 \left({u\over
2}\right) du\over (r + u^2)^{3/2}}
\label{3}
\eeq

From (\ref{3}), one can derive a certain number of properties of
$\omega$:
\beq
\omega > 0\quad\quad{\rm and} \quad\quad \lim_{r\rightarrow\infty}
\omega = 0~,
\label{4}
\eeq
\beq
{d\omega\over dr} < 0 ~,
\label{5}
\eeq
\vspace*{0.3cm}
\beq
{d\over dr} \left\vert {d\omega\over dr}/\omega\right\vert < 0~.
\label{6}
\eeq

Equations (\ref{4}) and (\ref{5})
are obvious. Inequality (\ref{6}) can be obtained by writing $\omega
{d^2\omega \over dr^2} -
({d\omega\over dr})^2$ as a symmetrized double integral following from
(\ref{3}).

In this section, we shall forget the origin of the problem and assume
only properties (\ref{4})
and (\ref{5}), and eventually (\ref{6}), but do not refer to the
specific expression (\ref{3}).
We shall write $\lambda\omega$ as $W$ and study the general properties
of the equation
\beq
h^{\prime\prime} = i~Wh~~.
\label{7}
\eeq
We shall sometimes write $h$ as
\beq
h = R \exp i~S~,
\label{8}
\eeq
$R$ and $S$ real.

Multiplying (\ref{7}) by $h^*$ and taking the real part of both sides we
get
\beq
h^* h^{\prime\prime} + 2h^{*\prime}h + h^{*\prime\prime}~h =
2h^{*\prime} h^\prime
\label{9}
\eeq
i.e.,
\beq
RR^{\prime\prime} = R^2~(S^{\prime})^2
\label{10}
\eeq
which implies that $R$ is $\underline{\rm convex}$, and since we are
interested in the solution
vanishing at infinity, $R$ is $\underline{\rm decreasing}$.

There, only the $\underline{\rm reality}$ of $W$ has been used. If we
take now the imaginary
part, we get
$$
h^*h^{\prime\prime} - hh^{*\prime\prime} = 2i W\vert h\vert^2~~,
$$
and integrating from $r$ to $\infty$:
\beq
R^2S^\prime = -\int^\infty_r~~W(r^\prime)R^2(r^\prime) dr^\prime
\label{11}
\eeq

Since $W$ is assumed to be $\underline{\rm positive}$ according to
(\ref{4}),
this means that the representation of the solution in the Argan diagram
is a spiral turning
clockwise and shrinking on the origin as $r$ goes from 0 to $\infty$.

Now we shall try to obtain inequalities relating $h$ and $h^\prime$.
First we multiply (\ref{7})
by $2h^\prime$ and integrate from $r$ to $\infty$. We get
\beq
-h^{\prime 2}(r) = -i~W~h^2(r) - i\int^\infty_r ~~{dW\over dr^\prime}
h^2(r^\prime) dr^\prime
\label{12}
\eeq

Since, From (\ref{5}) $dW/dr$ is negative, we can apply Wereistrass's
mean value theorem \cite{cc}
in the complex domain:
\beq
\int^\infty_r {dW\over dr^\prime} h^2(r^\prime) dr^\prime =
\overline{h^2} \int^\infty_r {dW\over
dr^\prime} dr^\prime = -\overline{h^2} W(r)~,
\label{13}
\eeq
where $\overline{h^2}$ is contained in the convex hull of
$h^2(r^\prime)$, and $r^\prime$
runs from $r$ to $\infty$. Since $\vert h\vert$  is decreasing, this
leads ot the following
inequality:
\beq
\vert h^\prime (r)\vert^2 < 2 W(r) \vert h(r)\vert^2~~.
\label{14}
\eeq
Except for a factor 2, this is what one would expect in a semi-classical
treatment.

A more subtle bound can be obtained  by multiplying (\ref{7}) by $h^*$
and
integrating from $r$ to $\infty$:
\beq
-h^*h^\prime = \int^\infty_r \vert h^\prime(r)\vert^2 dr^\prime + i
\int^\infty_r
W(r^\prime)\vert h(r^\prime)\vert^2 dr^\prime~.
\label{15}
\eeq
Hence,
$$
\bigg\vert \int^\infty_r {dW\over dr}(r^\prime) h^2(r^\prime) dr^\prime
\bigg\vert
< {{\rm sup}\atop {r<r^\prime < \infty}}  \bigg\vert
{{dW\over dr^\prime}(r^\prime) \over W(r^\prime)}
\bigg\vert \times \vert h \vert~~\vert h^\prime\vert~,
$$
and if property (\ref{6}) holds:
\beq
\bigg\vert \int^\infty_r {dW\over dr}(r^\prime) h^2(r^\prime) dr^\prime
\bigg\vert
< \bigg\vert{{dW\over dr}\over W}\bigg\vert \times {\vert h^\prime
\vert^2 + W\vert h
\vert^2\over 2\sqrt{W}}~.
\label{16}
\eeq

Inserting in Eq. (\ref{12}), we get
\beq
\vert \rho^2 - i\vert < {\vert{dW\over dr}\vert\over 2\vert W
\vert^{3/2}}~~\left[ 1 +
\rho^2\right]~,
\label{17}
\eeq
with
\beq
\rho = {h^\prime (r)\over \sqrt{W(r)} h(r)}
\label{18}
\eeq

From (\ref{17}), we get
$$
\left(\rho - {1+i\over \sqrt{2}}\right)~~\left(\rho + {1+i\over
\sqrt{2}}\right) \bigg\vert <
{\vert{dW\over dr}\vert / \vert W\vert^{3/2}\over 1-{1\over
2}\vert{dW\over dr}\vert / \vert
W\vert^{3/2}}
$$
provided ${1\over 2}~{dW\over dr}/\vert W\vert^{3/2} < 1$, and, noticing
that Re${h^\prime\over h}
< 0$ since $\vert h\vert$ is decreasing,
\beq
\bigg\vert\rho + {1+i\over\sqrt{2}}\bigg\vert < {\sqrt{2}\vert{dW\over
dr}\vert/\vert
W\vert^{3/2} \over 1 = {1\over 2} \vert{dW\over dr}\vert / \vert
W\vert^{3/2}}
\label{19}
\eeq

\section{Asymptotic behaviour of $h^\prime(0)/h(0)$}

We return now to our original problem i.e., we take $W = \lambda\omega$
with $\omega$ given
by Eqs. (\ref{2}) and (\ref{3}). What matters is the small $r$ behaviour
of $\omega$. For
$r\rightarrow 0$, we have
\beq
\omega \simeq {1\over 4} \ln {1\over r}~.
\label{20}
\eeq
In fact, over the whole range of $r$, we have found, numerically, that
\beq
\omega (r) \simeq {1\over 4} \ln {4+r\over r}
\label{21}
\eeq
with an error of at most 10\%.

The strategy is to estimate $h^\prime /h$ for some value of $r$
sufficiently close to zero, but
$\underline{\rm not~zero}$ because of the singularity of $\omega$, and
to estimate the error made. Without specifying $r$ yet, and only
assuming it is small we have
\beq
\bigg\vert {h^\prime (0)\over \sqrt{W(r)}} + {1+i\over\sqrt{2}}
h(0)\bigg\vert <
\bigg\vert {h^\prime (r)\over \sqrt{W(r)}
} + {(1+i)\over \sqrt{2}} h(r) \bigg\vert
+ {1\over\sqrt{W(r)}} \vert h^\prime (r) - h^\prime (0)\vert + \vert
h(r) - h(0)\vert~.
\label{22}
\eeq

From Eq. (\ref{7}), we obtain:
\beq
\vert h^\prime (r) - h^\prime (0)\vert = \bigg\vert\int^r_0 W
~h~dr^\prime\bigg\vert <
\vert h(0)\vert \int^r_0 W(r^\prime ) dr^\prime  ~~.
\label{23}
\eeq
Integrating (\ref{14}) from 0 to $r$ gives
$$
\bigg\vert\ln{h(r)\over h(0)}\bigg\vert < \int^r_0 \sqrt{W(r^\prime
)}dr^\prime~~,
$$
and, since, $\vert h(r)\vert$ decreases with $r$:
\beq
\vert h(r) -h(0)\vert < \vert h(0)\vert \int^r_0 \sqrt{W(r^\prime)}
dr^\prime~,
\label{24}
\eeq
and using again the fact that $W$ decreases, we get, combining
(\ref{18}), (\ref{21}), (\ref{22})
and (\ref{23})
\beq
\bigg\vert{h^\prime (0)\over \sqrt{W(r)}} + {1+i\over\sqrt{2}}~ h(0)
\bigg\vert < \vert h(0)\vert~~
\left[ {\sqrt{2}\bigg\vert{dW\over dr}\bigg\vert /W^{3/2}\over 1 -
{1\over 2} \bigg\vert {dW\over
dr}\bigg\vert /W} + {2\over\sqrt{W}} \int^r_0 W(r^\prime) dr^\prime
\right]~~,
\label{25}
\eeq
with the restriction $\vert{dW\over dr}\vert/W<1$. The problem is now to
optimise $r$ knowing
that $W\simeq {\lambda\over 4} \ln \left({1\over r}\right)$ for
$r\rightarrow 0$ (notice that
the large $r$ behaviour of $W$ is completely irrelevant for this
problem, as long as
$W\rightarrow 0$).

It is not very difficult to see that the qualitative optimum is reached
for
\beq
r = \lambda^{-1/2} (\log \lambda)^{-1}
\label{26}
\eeq
and gives
\beq
\bigg\vert {4\over \sqrt{\lambda\log\lambda}}~~{h^\prime(0)\over h(0)} +
1+i~\bigg\vert <
C(\log\lambda)^{-1/2}~,
\label{27}
\eeq
which is the final result of the article. This agrees with the result of
Ref. \cite{bb} (Ref.
\cite{aaa} contained an error).

	Whether the right-hand side of (\ref{27}) gives the order of magnitude
of the next term in the
asymptotic expansion of $h^\prime(0)/h(0)$ or not is unclear at the
present time. Inequality
(\ref{16}) is somewhat crude, because it disregards the phase changes in
the integral in the
left-and side. A much more careful analysis is needed. It is difficult
but not impossible and is
postponed to a future publication.

\vspace*{1cm}
\noindent
{\bf Acknowledgements}

We would like to thank S. Peign\'e and D. Schiff for submitting this
problem to us and for
stimulating discussions. One of us (A.M.) acknowledges the hospitality
of the LPTHPE, Orsay,
where his visit was supported by the CEC Science project SCI-CT91-0729.

\end{document}